\documentclass[pra, preprint]{revtex4}
\begin{document}
\title{Cavity broadcasting via Raman scattering}
\author{Moorad Alexanian}
\affiliation{\emph{Department of Physics and Physical
Oceanography, University of North Carolina at Wilmington,
Wilmington, NC 28403, USA, alexanian@uncw.edu}}
\date{\today}

\begin{abstract}

\textbf{Abstract.} The notion of broadcasting is extended to
include the case where an arbitrary input density state of a
two-mode radiation field gives rise to an output state with
identical marginal states for the respective modes, albeit
different from the input state. The initial unknown input density
state is unitarily related to the output state but is not equal to
the two identical output marginal states. This extended notion of
broadcasting suggests a possible way of discriminating between two
noncommuting quantum states.
\end{abstract}
\maketitle {}

\textbf{1. Introduction}

The no-cloning theorem for pure states, discovered in the early
1980s, is one of the earliest results of quantum computation and
quantum information. The no-cloning theorem states that it is
impossible to make a copy of an unknown, pure quantum state
\cite{WZ82}. The theorem places a limit on the ability to
manipulate quantum information. It follows from the no-cloning
theorem that no copying machine can make perfect copies of all
incoming states, if they are not all orthogonal to each other.
However, the whole idea of quantum cloning is to produce at the
output of the cloning machine two identical pure states, which are
unitarily connected with the input pure state. This less
restrictive condition for cloning was recently realized in the
exact cloning of nonorthogonal coherent states in a two-mode
cavity via Raman scattering \cite{MA03}.

The no-cloning theorem for pure states has been generalized and
extended to the case of mixed states \cite{BCF96}. The notion of
broadcast is introduced to include the possibility that although
an arbitrary mixed state may not be cloned nevertheless it may be
marginally reproduced. Thus cloning of mixed states would
represent a strong form of broadcasting. It should be remarked
that the notion of cloning or broadcasting considered \cite{BCF96}
refers to the case where the cloned or broadcasted output states
are precisely the same as the input state to be cloned or
broadcasted.\\

\textbf{2. Raman cavity dynamics for pure states}

In Ref. [2], the dynamics of a two-level atom, singly injected
into a two-mode, high-$Q$ cavity was studied. The atom-fields
interaction is given by the Raman-coupled Hamiltonian $H_{int}$.
The time evolution of an arbitrary state of the atom-fields system
is determined in the interaction picture by
$|\Psi(t)>=U(t)|\Psi>$, where $U(t)=e^{-iH_{int}t/\hbar}$. Steady
states can be achieved inside the cavity by singly injecting atoms
that are in a coherent superposition of their upper and lower
states, where the admixture of the atomic states $|1>$ and $|2>$
represents the tunable parameter of the experiment. The successive
iterations give the following reduced fields density matrix
$\rho_{l}$ after $l$ such atoms have singly traversed the cavity
\begin{equation}
\rho_{l}=\textup{tr}_{a}[U(\tau)\rho_{l-1}\rho_{a}U^{\dag}(\tau)],
\end{equation}
where the trace is over the atomic states
$\rho_{a}=(\alpha|1>+\beta|2>)(<1|\alpha^*+<2|\beta^*)$, which is
the state of the injected atoms, and $\tau$ is the interaction
time. If the limit of the iterations exists, then the resulting
state is called a steady state, which is a fixed point of the map
(1).

The state
\begin{equation}
|\Psi_{steady}>=\sum_{n_{1},n_{2}=0}^{\infty}C_{n_{1},n_{2}}|n_{1},n_{2}>
\otimes \hspace{.1in}(\alpha|1>+\beta|2>),
\end{equation}
where
\begin{equation}
C_{n_{1},n_{2}}=\left(-\frac{\alpha}{\beta r}\right)^{n_{2}}
\hspace{0.05in}\sqrt{\frac{(n_{1}+n_{2})!}{n_{1}!
\hspace{0.05in}n_{2}!}} \hspace{0.05in} C_{n_{1}+n_{2},0},
\end{equation}
is a fixed point of the map (1) and so constitutes a steady state
\cite{MA03}.

The dynamics governed by the Raman Hamiltonian conserves the total
number of photons in the two modes. Accordingly, if the cavity
fields in the two modes are initially disentangled with field mode
one in a coherent state $|\gamma>$ and field mode two in the
vacuum state $|0>$, then the steady state of the electromagnetic
fields inside the cavity is a two-mode coherent state and the
effect of atom-fields interactions inside the cavity is to give
rise to the following relationship between the initial input state
and the output steady state \cite{MA03}
\begin{equation}
|\gamma>_{1} \otimes \hspace{.1in}|0>_{2} \longrightarrow \left|
\frac{\gamma}{\sqrt{1+|\frac{\alpha}{\beta r}|^2}}\right>_{1}
\otimes \hspace{.1in} \left|\frac{-\frac{\alpha}{\beta
r}\gamma}{\sqrt{1+|\frac{\alpha}{\beta r}|^2}}\right>_{2} ,
\end{equation}
where the subscripts indicate the modes of the radiation fields
and $r$ is the ration of the atom-photon coupling constants.

The cavity transformation (4) is effected by the unitary, quantum
beam splitter operator
\begin{equation}
S(\lambda) =e^{\lambda
a^{\dag}_{1}a_{2}-\lambda^{*}a^{\dag}_{2}a_{1}},
\end{equation}
where $\lambda=|\theta|e^{-i\phi}$ with $\alpha/\beta
r=e^{i\phi}\textup{tan}|\theta|$ and $a^{\dag}_{i}$ ($a_{i}$),
$i=1,2$, represents the creation (annihilation) operators for the
respective fields. The cloning of the coherent state $|\gamma>$ is
achieved with input state $|\sqrt{2}\gamma>_{1}\otimes
\hspace{.05in} |0>_{2}$ when the tuning parameter $\alpha/\beta
r=-1$. This value for the tuning parameter corresponds to a 50/50
quantum beam splitter. There is an important distinction between
an optical 50/50 beam splitter and the quantum beam splitter
generated by the cavity transformation (4).  Both beam splitters,
effected by operator (5), conserve the total number of photons
between the input and output states. The optical beam splitter
conserves energy owing to the equality of the input and output
photon frequencies. However, in the cavity transformation (4),
modes one and two have, in general, different frequencies and so
one must consider the energy gained or lost by the atoms
traversing the cavity in order to conserve the total energy.
Transformation (4) resembles a parametric down-conversion whereby
the fraction $|\alpha \beta^{-1} r^{-1}|^2/(1+ |\alpha \beta^{-1}
r^{-1}|^{2})$ of photons in the input coherent state with the
higher frequency $\omega_{1}$ are converted into photons in the
output coherent state with the lower frequency $\omega_{2}$. Total
energy conservation is established if the energy of the traversing
atoms are taken into account since the traversing atoms make
transitions from the atomic ground state $|1>$ to the atomic
excited state $|2>$ and vice versa via Raman scattering. Optimal
universal cloning can be realized using parametric down-conversion
and also via stimulated emission \cite{SWZ00}. Our quantum beam
splitter also resembles a ``parametric up-conversion" when the
initial input state is given by $|0>_{1}\otimes
\hspace{.1in}|\gamma>_{2}$. The fraction $1/(1+ |\alpha \beta^{-1}
r^{-1}|^{2})$  of photons in the input state at the lower
frequency $\omega_{2}$ are converted in the output state into
photons with the higher frequency $\omega_{1}$.\\

\textbf{3. Raman cavity dynamics for mixed states}

Consider the general case where initially inside the cavity mode
one is in an arbitrary mixed state while mode two is in the vacuum
state. One can write the initial state $\rho_{i}$ inside the
cavity in the coherent state representation and so
\begin{equation}
\rho_{i}=\int d^2\gamma \hspace{.06in}
P(\gamma,\gamma^*)|\gamma>_{1}<\gamma|\hspace{.04in}\otimes\hspace{.04in}|0>_{2}<0|,
\end{equation}
where $P(\gamma,\gamma^*)$ is some real function of $\gamma$ and
$\gamma^*$ which is normalized to unity and the subscripts
indicate the modes of the radiation field.

The resulting steady state or equivalently the state
$\rho_{s}=S(\lambda)\rho_{i}S^\dag(\lambda)$ is
\[
\rho_{s}= \int d^2\gamma \hspace{.06in} P(\gamma,\gamma^*)
\hspace{.1in}\left | \frac{\gamma}{\sqrt{1+|\frac{\alpha}{\beta
r}|^2}}\right>_{1}\left<
\frac{\gamma}{\sqrt{1+|\frac{\alpha}{\beta r}|^2}}\right|
\hspace{.05in}\]
\begin{equation}
\otimes\hspace{.05in} \left|\frac{-\frac{\alpha}{\beta
r}\gamma}{\sqrt{1+|\frac{\alpha}{\beta r}|^2}}\right>_{2}
\left<\frac{-\frac{\alpha}{\beta
r}\gamma}{\sqrt{1+|\frac{\alpha}{\beta r}|^2}}\right|,
\end{equation}
where the corresponding marginal density operators for modes one
and two are given by
\begin{equation}
\rho_{1}=\int d^2\gamma \hspace{.06in}
P(\gamma,\gamma^*)\hspace{.03in}\left|\frac{\gamma}{\sqrt{1+|\frac{\alpha}{\beta
r}|^2}}\right> \left <\frac{\gamma}{\sqrt{1+|\frac{\alpha}{\beta
r}|^2}}\right|
\end{equation}
and
\begin{equation} \rho_{2}= \int d^2\gamma \hspace{.06in}
P(\gamma,\gamma^*) \hspace{.1in} \hspace{.1in}
\left|\frac{-\frac{\alpha}{\beta
r}\gamma}{\sqrt{1+|\frac{\alpha}{\beta r}|^2}}\right>
\left<\frac{-\frac{\alpha}{\beta
r}\gamma}{\sqrt{1+|\frac{\alpha}{\beta r}|^2}}\right|,
\end{equation}
respectively.  If the tuning parameter $\alpha/\beta r=-1$, then
the arbitrary input state $\int d^2\gamma \hspace{.06in}
P(\gamma,\gamma^*)|\gamma>\hspace{.053in} <\gamma|$ has an output
state where both modes of the field are in the same, precise
marginal state given by
\begin{equation}
\rho_{b}= \int d^2\gamma \hspace{.06in}
P(\gamma,\gamma^*)|\frac{\gamma}{\sqrt{2}}>
<\frac{\gamma}{\sqrt{2}}|.
\end{equation}
This result indicates that arbitrary states may be broadcasted,
where the reduced density matrices of the two field modes of the
output state are identical but differ from the input state and are
unitarily connected to it. Therefore, if the arbitrary state $\int
d^2 \gamma \hspace{.06in}
P(\gamma,\gamma^*)|\gamma>\hspace{.053in} <\gamma|$ is to be
broadcasted, then the cavity must have one mode initially in the
vacuum state and the other in the state $\frac{1}{2}\int d^2
\gamma \hspace{.06in}
P(\frac{\gamma}{\sqrt{2}},\frac{\gamma^*}{\sqrt{2}})|\gamma>\hspace{.053in}
<\gamma|$. Note, however, that the input state
$|\sqrt{2}\gamma>_{1}\otimes \hspace{.05in}|0>_{2}$ is the only
initial state \cite{MA03} that can lead to a disentangled pure
output state, viz.
$|\gamma>_{1}\otimes\hspace{.05in}|\gamma>_{2}$, and thus to the
exact cloning of $|\gamma>$. This is achieved for
$P(\gamma',\gamma'^{*})=\delta^{2}(\gamma'-\sqrt{2}\gamma)$.

It is important to remark that in a Gaussian quantum-cloning
machine \cite{CIR00}, which can be viewed as the continuous
counterpart of the universal qubit cloner, one has to prepare an
initial coherent state $|\alpha>$ as input and its two outputs are
a mixture of coherent states characterized by a density matrix
with fidelity 2/3.  In the present work, instead, one has to
prepare the initial coherent state $|\sqrt{2}\alpha>_{1}$ in order
to produce the pure output state $|\alpha>_{1} \otimes
\hspace{.1in}|\alpha>_{2}$. The universal cloner gives rise to a
mixed output state in the cloning of a coherent state while the
present cloner gives rise to two identical coherent states as
output. Our cloning machine is a ``universal cloning machine" for
coherent states since the cloning process is input independent.

The mean electric-field strength in a coherent state $|\xi>$ looks
like the electric-field strength of a coherent, classical
radiation with mode amplitude $\xi$. In addition, the relative
noise of the electric-field strength in the coherent state $|\xi>$
is inversely proportional to $|\xi|$ and so the relative noise
decreases for increasing mean photon number. Accordingly, coherent
states $|\xi>$ with larger mean photon numbers are closer to their
classical, coherent wave counterpart. The laser field well above
threshold is described as being in a coherent state. The mean
laser intensity in the steady-state grows in proportion to the
pump parameter while the relative mean squared light intensity
fluctuation goes down with increasing value of the pump parameter.
The quantum state of the laser field can be close to a coherent
state, where the photon occupation number can be exceedingly large
and the radiation produced by the laser comes close to being
classical. Accordingly, coherent states $|\xi>$ with large mean
photon numbers are closer to their classical, coherent wave
counterpart and thus are easier to produce. A weak coherent state
can be obtained in turn from a more intense coherent state
\cite{SBRF93}.

Note that Eq. (6) need not represent a classically correlated
density matrix \cite{RFW89}. The real function
$P(\gamma,\gamma^{*})$ may assume negative values and be more
singular than a delta function and thus possess no classical
analog. Note also that even if $P(\gamma,\gamma^{*})$ behaved like
a true probability density, i.e., it is non-negative and not more
singular than a delta function, it would still not describe
probabilities owing to the non-orthogonality of the coherent
states.\\

\textbf{4. Non-orthogonal mixed state discrimination}

The results presented here also suggest a possible way to achieve
a probabilistic error-free discrimination, that is, one which
sometimes fails, but when successful never gives an erroneous
result. This procedure is referred to as an unambiguous
discrimination \cite{RST03}. If the initial input state is the
arbitrary state (6), then the output reduced density matrix of
mode one is obtained from (7) and so
\begin{equation}
\rho_{A}=\int d^2\gamma \hspace{.06in} P(\gamma,\gamma^*)|A
\gamma>_{1}<A \gamma|,
\end{equation}
where
\begin{equation}
A = \frac{1}{\sqrt{1+|\frac{\alpha}{\beta r}|^2}}.
\end{equation}
Similarly, for the initial state
\begin{equation}
\sigma_{i}=\int d^2\lambda \hspace{.06in}
Q(\lambda,\lambda^*)|\lambda>_{1}<\lambda|\hspace{.04in}\otimes\hspace{.04in}|0>_{2}<0|,
\end{equation}
with output reduced density matrix
\begin{equation}
\sigma_{A}=\int d^2\lambda \hspace{.06in} Q(\lambda,\lambda^*)|A
\gamma>_{1}<A \gamma|.
\end{equation}
In the limit $|\frac{\alpha}{\beta r}|\rightarrow \infty$, i.e,
when the input atoms are essentially in the atomic ground state
$|1>$, $A\rightarrow 0$ and so the contributions to $\rho_{A}$ and
$\sigma_{A}$ come mainly from the vacuum $|0>$ and the one photon
state $|1>$ hence
\[\rho_{A}=\left[1- A^2<|\gamma|^2>\right]|0><0|+ A <\gamma^*>|0><1|\]
\begin{equation}
+ A<\gamma>|1><0|+ A^2<|\gamma|^2>|1><1|,
\end{equation}
where $<|\gamma|^2>=\textup{tr}(\rho_{i} a^\dag_{1}a_{1})$ and
$<\gamma> =\textup{tr}(\rho_{i} a_{1})$ for the initial state
$\rho_{i}$. Similar equations hold for $\sigma_{A}$ where the
expectation values are, instead, in terms of $\sigma_{i}$. The
diagonal terms in (15) are given by
\begin{equation}
<n|\rho_{A}|n> =\frac{A^{2n}}{n!}
\int|\gamma|^{2n}e^{-A^2|\gamma|^2}\hspace{.1in}
P(\gamma,\gamma^*)d^2\gamma
\end{equation}
and so the many-photon contributions in (15) are vanishingly small
for arbitrarily small values of $A$ and such terms and their
corresponding off-diagonal matrix elements have been neglected in
(15).

For arbitrary $<\gamma>$ and $<|\gamma|^2>$, the density matrix
(15) represents a mixed state. Therefore, the supports of
$\rho_{A}$ and $\sigma_{A}$ are equal and so, in general, the two
states cannot be discriminated \cite{RST03}. However, if
\begin{equation}
<|\gamma|^2> - |<\gamma>|^2 = A^2(<|\gamma|^2>)^2 \hspace{.1in}\ll
\hspace{.1in} <|\gamma|^2>,
\end{equation}
then $\rho_{A}=|\Phi_{\rho}><\Phi_{\rho}|$ is a pure state with
\begin{equation}
|\Phi_{\rho}>=[1-A^2<|\gamma|^2>]^{1/2}\hspace{.1in}\left[|0>+
A\frac{<|\gamma|^2>}{<\gamma^{*}>}|1>\right].
\end{equation}
Similar results hold for $\sigma_{A}$ with
$\sigma_{A}=|\Phi_{\sigma}><\Phi_{\sigma}|$ and $|\Phi_{\sigma}>$
is given by (18) with expectation values taken with respect to
$\sigma_{i}$ rather than $\rho_{i}$. One may perform measurements
in order to optimally distinguish \cite{IDP87} between the two
nonorthogonal states $|\Phi_{\rho}>$ and $|\Phi_{\sigma}>$. Such
measurements are described by a positive operator-valued measure
(POVM) \{$E_{\rho}, E_{\sigma}, E_{?}$\} \cite{DK76}, where
$E_{\rho} = C_{\rho}|\Phi^{orth}_{\sigma}><\Phi^{orth}_{\sigma}|$,
$E_{\sigma} = C_{\sigma}|\Phi^{orth}_{\rho}><\Phi^{orth}_{\rho}|$,
$E_{?}=1-E_{\rho}-E_{\sigma}$, and
$<\Phi_{\rho}|\Phi^{orth}_{\rho}>=
<\Phi_{\sigma}|\Phi^{orth}_{\sigma}> =0$.  Positive constants
$C_{\rho}$ and $C_{\sigma}$ can always be found such that $E_{?}$
is a positive operator. The measurement procedure can have up to
three possible outcomes, associated with identifying the state
$\rho_{i}$, identifying the state $\sigma_{i}$, and failing to
identify the state conclusively \cite{RST03}. The discrimination
between the mixed states $\rho_{i}$ and $\sigma_{i}$ is optimized
for appropriate values of $C_{\rho}$ and $C_{\sigma}$ consistent
with the positivity of the operator $E_{?}$. The successful
unambiguous discrimination \cite{RST03,IDP87} gives the maximum
probability $P^{max}= 1- |<\Phi_{\sigma}|\Phi_{\rho}>|$ for the
case of equal \emph{a priori} probabilities. A practical proposal
for preparing chosen superpositions of the vacuum and the
one-photon states \cite{PPB98} may be used for the above POVM.
Also, optimal unambiguous weak coherent states discrimination may
be realized using polarization beam splitters \cite{HIGM95}. Weak
coherent states are easy to generate from strong coherent states
\cite{SBRF93}. In an effort to increase the maximum unambiguous
discrimination between the attenuated states $|\Phi_{\rho}>$ and
$|\Phi_{\sigma}>$, one can consider weaker attenuation and thus
include higher photon number states in (15). This would require
optimum measurements for three or more distinct outcomes
\cite{CKCBRS01}.

The above approach to discriminate between non-orthogonal mixed
states represents an optical attenuation of an arbitrary mixed
states into the vacuum and the one-photon subspace. Note that
$P^{max} = O(A^2)$ and so the success probability for unambiguous
discrimination would be rather small but increases significantly
for mixed quantum states with small mean photon numbers. Our
procedure does not represent a ``quantum scissors" for density
matrices since the vacuum and the one-photon matrix elements of
the attenuated density matrix differs from the vacuum and the
one-photon matrix elements of the original density operator.
Quantum entanglement and the nonlocality of a single photon has
been used to truncate the number state expansion of a pure optical
state thus leaving only the vacuum and one-photon components
\cite{PPB98}. However, at present there is no known scheme to
project an arbitrary density matrix onto the zero- and one-photon
subspace.

The class of mixed states that yield to unambiguous discrimination
are determined by (17) and are close to a coherent state and
includes, for instance, superposition of two or more coherent
states. Of course, the procedure of identifying mixed states, by
reducing the support of the mixed states $\rho_{i}$ and
$\sigma_{i}$ to the vacuum and the one photon state, was
considered for states for which $\textup{tr}(\rho a)=<\gamma>\neq
0$. Mixed classical states such as the randomly phased laser
model, thermal light, etc., where $P(\gamma,\gamma^*)=
P(|\gamma|)\geq 0$ and so $<\gamma>= 0$, do not yield to an
unambiguous discrimination since for these cases the states
$\rho_{A}$ and $\sigma_{A}$ commute and so the original mixed
states $\rho_{i}$ and $\sigma_{i}$ cannot be discriminated. The
latter is expected since such type of mixed states are diagonal in
the number of particle representation with non-zero diagonal
matrix elements and thus possess the same support.\\

\textbf{5. Summary and Conclusion}

In closing, the no-cloning theorem \cite{WZ82} and the
no-broadcasting theorem \cite{BCF96} impose fundamental quantum
mechanical restrictions on the ability to copy or broadcast
arbitrary states. Both theorems are combined into one by proving
that noncommuting mixed states cannot be broadcast and that
cloning represents a strong form of broadcasting \cite{BCF96}. In
Ref. [2], the cloning of coherent states by a cavity-cloning
machine suggests a weaker operational definition of cloning or
copying. In the former, the cloning machine produces two identical
pure states as output \cite{MA03} that are unitarily related to a
different input state. In the latter the input and the two output
states are required to be identical \cite{WZ82}.

In this work, the meaning of broadcasting \cite{BCF96} is weakened
to mean that the output density state for the two-mode radiation
field are in the same marginal state, which differs from the input
state but is obtained from it by a unitary transformation. Such
more modest, operational definition of quantum copier has also
been considered previously \cite{BVP97}. The ability of the
unambiguous discrimination of two mixed states ought to be of
interest in quantum cryptography where the usual procedure is to
encode information into noncommuting mixed states in order to
prevent eavesdropping.

\bibliography{basename of .bib file}

\end{document}